\documentclass{aa}

\def\lxlbol{$\log \mathrm{L}_\mathrm{X}/\mathrm{L}_\mathrm{bol}$}
\usepackage{graphicx}
\usepackage{aalongtable}
\usepackage{natbib}
\bibpunct{(}{)}{;}{a}{}{,}%
\usepackage{txfonts}
\begin{document}
   \title{Deep X-ray survey of the young open cluster NGC 2516 with XMM-Newton.
\thanks{Based on observations obtained with XMM-Newton, an ESA science mission 
with instruments and contributions directly funded by ESA Member States and 
NASA.}}


\author{I. Pillitteri\inst{1} \and
        G. Micela\inst{2} \and
        F. Damiani\inst{2} \and
        S. Sciortino\inst{2}
       }

\offprints{I. Pillitteri}

\institute{Universit\`a degli Studi di Palermo,
Piazza del Parlamento 1, 90134, Palermo -- ITALY --\\
\email{pilli@astropa.unipa.it}
\and
INAF - Osservatorio Astronomico di Palermo
Piazza del Parlamento 1, 90134 Palermo - ITALY - \\
\email{giusi@astropa.unipa.it,damiani@astropa.unipa.it,sciorti@astropa.unipa.it}
}
\date{Received; accepted }

\abstract
{}
{We report on a deep X-ray survey of the young ( $\sim$ 140 Myr), rich
open cluster NGC 2516 obtained with the EPIC camera on board the XMM-Newton
satellite.} 
{By combining data from six observations, a high sensitivity, greater
than a factor 5 with respect to recent Chandra observations, has been achieved.
Kaplan-Meier estimators of the cumulative X-ray luminosity distribution are 
built, statistically corrected for non members contaminants and compared 
to those of the nearly coeval Pleiades.
The EPIC spectra of the X-ray brightest stars are fitted using optically 
thin model plasma with one or two thermal components.}
{We detected 431 X-ray sources and 234 of them have as optical counterparts
cluster stars spanning the entire NGC 2516 Main Sequence. 
On the basis of X-ray emission and optical photometry, we indicate 20 new 
candidate members of the cluster; 
at the same time we find 49 X-ray sources without
known optical or infrared counterpart.   The X-ray 
luminosities of cluster stars span the range $\log \mathrm L_\mathrm X$ 
(erg s$^{-1}$) $=$ 28.4 -- 30.8.
The representative temperatures span the 0.3 -- 0.6 keV (3.5 -- 8 MK) 
range for the cool component and 1.0 - 2.0 keV
(12 -- 23 MK) for the hot one; similar values are found in other young
open clusters like the Pleiades, IC 2391, and Blanco 1. While no significant 
differences are found in X-ray spectra, NGC 2516 solar type stars are definitely
less luminous in X-rays than the nearly coeval Pleiades.
The comparison with a previous
ROSAT survey evidence the lack of variability amplitudes larger than a factor 2
in solar type stars in a $\sim 11$ yr time scale of the cluster and thus
activity cycles like in the Sun are probably absent or different by period and 
amplitude in young stars.}
{}

\titlerunning{Deep X-ray survey of NGC 2516 with XMM-Newton.}
\authorrunning{Pillitteri et al.}
   \maketitle
%
\section{Introduction}
X-ray observations of stars have shown that
X-ray emission is a common feature along the main sequence 
\citep{Vaiana1981}. The emission mechanisms depends on masses: in massive
O and early B type stars the X-rays are likely generated by plasma shocked 
and heated in stellar winds, although a role of magnetic field is 
not ruled out \citep{Lucy1980,Feldmeier95,Babel97,Cassinelli2000,Waldron01}, 
whereas solar mass stars present coronae composed by hot,  magnetically 
confined emitting plasma, as directly observed in the Sun \citep{Vaiana1981}.
Only B-late -- A-early stars are not expected to emit significant X-rays 
because neither strong stellar winds nor convective regions, necessary 
to have a solar magnetic dynamo, are present in these stars.
Focusing on solar mass stars, the X-ray activity is strictly 
linked to the internal structure and the rotation. The coupling of 
convective motions and rotation produces the solar-type dynamo. 
The losses of angular momentum is the main factor of decrease of X-ray activity
during stellar evolution, especially in the first Gyr, although the role of 
other factors like, e.g.,
stellar chemical composition or environmental forming conditions cannot
be ruled out. 

In this context the studies based on stellar open clusters with different
properties are of great importance to better define the relevance of
age differences, chemical composition and environment influence.  
An open cluster like the Pleiades constitutes a prototype  
for the study of activity and evolution at an age around 100 Myr, 
given its richness and nearby distance. 
The evolution of X-ray emission with stellar age has been studied 
in the last five years with XMM-Newton and Chandra X-ray observatories, 
which provided high quality surveys of open clusters like Pleiades, 
Blanco 1, IC 2391, NGC 6383, NGC 188, M67 
(cf. \citealp{Briggs2003}, \citealp{Pilli2004},
\citealp{Marino04}, \citealp{Rauw2003}, \citealp{vanBerg04} and 
\citealp{Gondoin05}).
In this paper we study the X-ray emission of the young open cluster NGC 2516.

NGC 2516 is a rich open cluster with an age of 140 Myr \citep{Meynet93}, 
slightly older than the Pleiades, located in the southern hemisphere at a
height of $\sim$100 pc on the Galactic Plane.
The richness of star and its relatively nearby 
distance (about 387 pc) have motivated the growing interest for the study of 
this cluster both in the optical and in the X-ray band.
In the optical band the  photometric study  of \citet{Dachs70} and 
\citet{DachsKab89} have shown a higher number of stars and a larger fraction 
of peculiar Ap stars with respect to the Pleiades.
Episodes of subsequent star formation, as
due to passages across the Galactic Plane or interaction with the Gum Nebula,
have been also speculated by \citet{DachsKab89} in
order to explain younger blue stragglers in this cluster. 
More recently, the deep photometric survey of \citet{Jeffries01} 
allowed to trace the cluster sequence down to V $\simeq$ 20 and
to build an optical catalog of 1254 photometric members.
The {\em Mass Function} of the cluster  is consistent with the empirical 
Salpeter's power 
law $dN/dM \propto M^{-1.35}$ but with index 1.47--1.67, 
depending on the assumed metallicity. The chemical
composition remains to date quite uncertain for this cluster: 
while photometry suggests a significant underabundance of metals 
([Fe/H] = -0.3, \citealp{Jeffries01}), more recently \citet{Terndrup02},
on the basis of high resolution spectroscopy of two stars,
strongly suggested a rather solar-like metallicity for NGC 2516. 
Furthermore, these authors conclude that an age of 150 Myr and a distance
modulus of 7.93 mag (i.e. a distance of $\sim$ 385.5 pc) are more appropriate.

In the X-ray band \citet{Jef97} and \citet{Micela00} analyzed ROSAT PSPC and HRI
observations leading to 47 detections among cluster stars with
spectral type from B2 down to K. The G- and K-type stars of the cluster 
appeared under-luminous when compared to the Pleiades thus 
inducing to consider the low metallicity of the cluster (estimated at that time 
by photometry) to have a role in determining the X-ray emission. 
With the advent of {\em Chandra} and {\em XMM-Newton} missions, the cluster 
has been studied by \citet{Harnden01}, \citet{Sciortino01} and 
\citet{Damiani2003}, confirming the under-luminosity of solar type stars 
with respect to the Pleiades and reporting X-ray emission from several 
Ap stars. 
In M-type stars the X-ray emission was found similar to the Pleiades.
The time X-ray variability of stars in NGC2516 has been studied by
\citet{Wolk04} with Chandra observations over a 2 year time span.

Unfortunately, cluster membership is determined only through 
photometry, since cluster proper motion is close to the solar value, making very
difficult to obtain reliable astrometric membership determination. Therefore
the low X-ray emission level in solar type stars could be attributed to both a 
significant contamination of the cluster sample or to a real effect, related 
to slightly older age of NGC~2516 with respect to the Pleiades.

In the present paper we report on a deep X-ray survey of NGC 2516 obtained through
a series of observations with XMM-Newton satellite. Our aim is to determine
with the highest sensitivity the X-ray emission of the cluster. 
In particular we will focus on
the X-ray properties of solar type stars of this cluster and their comparison
with those of the very similar and well studied Pleiades.
The structure of the paper is as follows: in Sect. 2 we present the data 
and the analysis method. In Sect. 3 we discuss the spectral properties and
the X-ray emission level of solar type stars. 
In Sect. 4 we compare the X-ray luminosities from XMM-Newton with those of
Chandra and ROSAT.
Sect. 5 reports a discussion on other stars suggested as possible new members 
of the cluster, based on X-ray emission and optical photometry.
In Sect. 6 we give a brief summary of relevant results.


\section{The observations and the data analysis}

\begin{figure*}
\centering
\includegraphics[height=\textwidth,angle=-90]{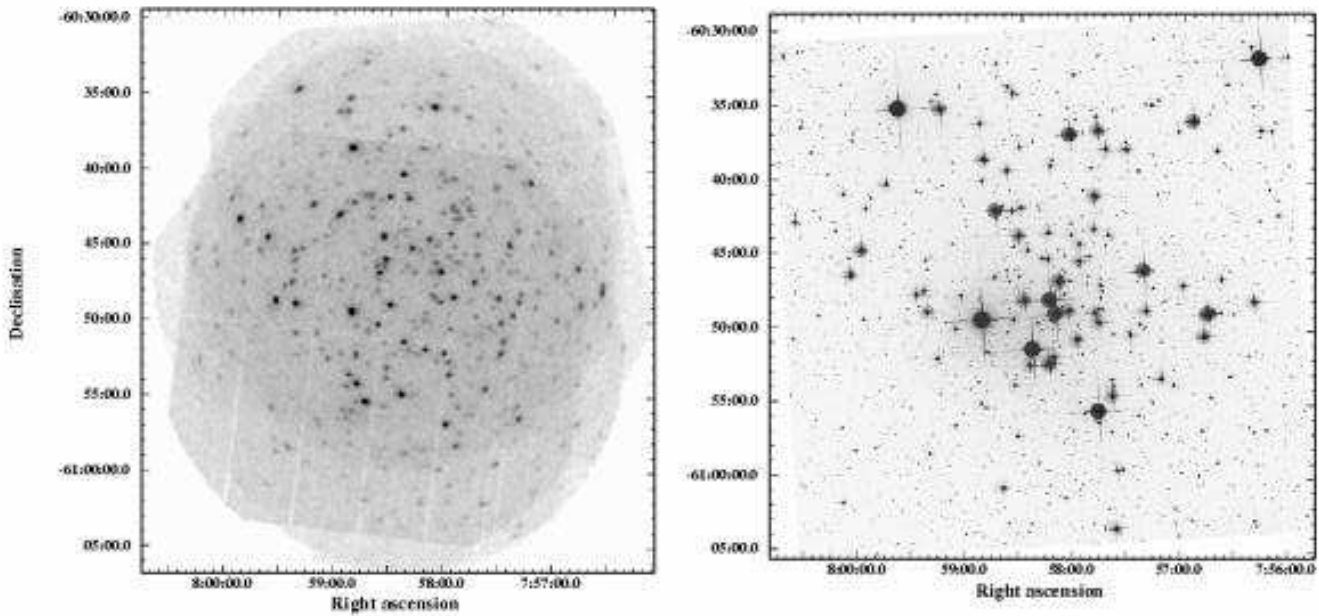}
\caption{\label{img} X-ray (left panel) and optical (right panel) 
images of the NGC 2516 cluster. The X-ray survey covers a sky region of
$\sim 18\arcmin \times 25\arcmin$ and contains 593 cluster stars (cf. Sect. 2.1)}%
\end{figure*}
NGC 2516 has been observed several times with XMM-Newton during the first 
two years of satellite operations for calibration purposes. 
Table \ref{obslog} summarizes the relevant information of the observations 
we have used.
They span a period of 19 months with exposure times between 10 and 20 ks. 
All of these observations have been performed with the thick filter. 
Fig. \ref{img} shows the summed EPIC exposures (left panel) and the 
optical image of the same region (right panel). 
\begin{table*}
\centering
\small
\caption{\label{obslog}Log of XMM-Newton observations of NGC 2516. All
exposures have been taken with the {\em Thick} filter.Positions refer to J2000,
times (referred to the pn exposures) are in kilo seconds.}
\vspace{0.3cm}
\begin{tabular}{l l l l l l}\hline \hline
OBS Id. & Orbit Nr. & R.A. & Dec. & Time (filtered/total) & Date \\ \hline
0113891001 & 0060 & 7:58:20 & -60:52:13 & 20.0/20.0 & 2000-04-06 @ 19:25:05\\
0113891101 & 0060 & 7:58:20 & -60:52:13 & 14.5/17.0 & 2000-04-07 @ 04:05:31\\
0126511201 & 0092 & 7:58:22 & -60:45:36 & 21.9/27.5 & 2000-06-10 @ 08:56:51\\
0134531201 & 0209 &7:58:22 & -60:45:36 & 18.8/19.0  & 2001-01-29 @ 00:23:20\\
0134531301 & 0209 &7:58:22 & -60:45:36 & 9.3/9.3 &    2001-01-29 @ 08:01:48\\
0134531501 & 0346 &7:58:22 & -60:45:36 & 21.2/21.2 &  2001-10-29 @ 09:53:59\\ \hline
\end{tabular}
\end{table*}

\begin{figure}
\begin{center}
\includegraphics[height=\columnwidth,angle=-90]{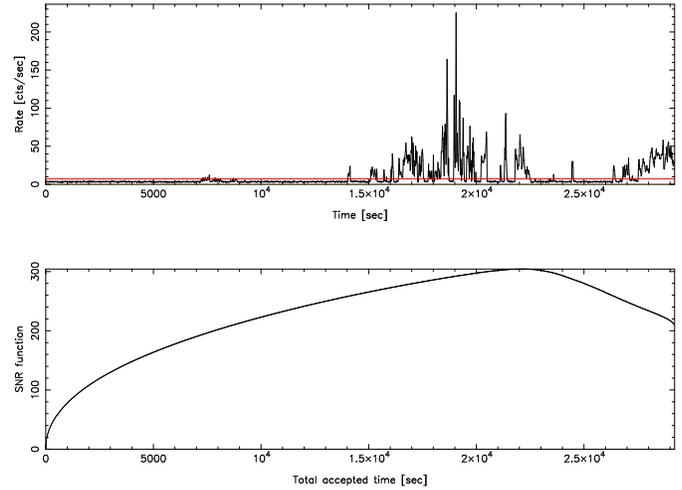}
\caption{\label{bkgrate} Light curve for MOS 1 of EPIC observation during 
the XMM orbit 0092 (top panel); 
the horizontal line is the threshold rate that
corresponds to filter the event times in order to maximize the S/N function,
plotted in the bottom panel as a function of the total accepted times.}
\end{center}
\end{figure}

The data of each EPIC detector (MOS 1, 2 and pn) have been reduced with SAS 
software in order to obtain tables of photons with calibrated 
astrometry, arrival times and energies. 
Subsequently we have chosen only events in EPIC
field of view, with energies in 0.3-7.9 keV band, and that have triggered 
up to four pixels simultaneously.
Furthermore, the data have been screened from 
high background rate intervals by maximizing the signal-to-noise ratio
of weak sources. 
Fig. \ref{bkgrate} shows an example of the time screening procedure for
EPIC pn of observation during the orbit 0092, the worst case in our set
of observations.
The light curve of all photons is plotted in the top panel. Spikes
are due to intervals of high background rate. 
Bottom panel shows the signal to noise ratio as a function of the accepted 
time. We have chosen a rate threshold that filters only the time intervals 
yielding the maximum signal to noise ratio in the filtered image
in order to optimize the detection of weakest sources.
The derived fraction of exposure for each observation (pn exposures only) 
is reported in Table \ref{obslog}. 

Source detection and X-ray photometry derived from the sum of all the MOS and 
pn exposures have been obtained with PWXDetect code, developed at  
INAF~-~Osservatorio Astronomico di Palermo, derived from the original
ROSAT code for source detection \citep{Dami97.1,Dami97.2}, and based 
on the analysis of the wavelet transform of the count rate image. 
It allows to combine data from different EPIC detectors also taken in different
observations in order to achieve the deepest sensitivity. 
The code provides estimates of the positions, total counts, rates,
effective exposure times and significance of each detection. 

We have chosen MOS 1 as a reference detector to evaluate rates, 
fluxes and luminosities. The scaling factor between MOS 1, MOS 2 and pn depends on
the source spectrum, the filter used during observation and the chosen
energy band. In order to properly scale the rates of pn to MOS1 
in reason of the different efficiency, we have
used the ratios of source count rates, as deduced by photons accumulated in a 
15$\arcsec$ radius circular region at the source positions. 
In this way we use directly the data to find the relative instrument
efficiencies without any hypotheses on source spectrum.
The rates in the summed image are equivalent to MOS 1 count rates (CR) 
according to the following formula:
\[\mathrm{CR}_\mathrm{sum} = \Sigma_\mathrm{MOS1} \mathrm{CR}_\mathrm{MOS1} + 
\Sigma_\mathrm{MOS2} \mathrm{CR}_\mathrm{MOS2}\cdot r_\mathrm{MOS2/MOS1} 
+ \Sigma\mathrm{CR}_\mathrm{pn}\cdot r_\mathrm{pn/MOS1} \]
where the medians of the scaling factors $r_\mathrm{MOS2/MOS1}$ and 
$\mathrm{pn/MOS1}$ are respectively $\sim$ 1 and $\sim$ 3.6, as evaluated from 
source count rates measured by different detectors.
The exposure times used to calculate CRs are obtained from exposure maps, 
and thus they take into account any spatial 
non-uniformities due to vignetting, RGS grating obscuration, chip geometry 
etc.

In the combined EPIC datasets we have detected 431 X-ray sources with a 
significance level greater than 5.0 $\sigma$, which should lead 
statistically to at most one spurious source in the field of view. 
The threshold has been calibrated  by applying the source detection algorithm
to 500 simulated datasets of only background photons.
In Table  \ref{xdet} we report the list of sources and their properties.

\subsection{X-rays from NGC 2516: fluxes and luminosities} 
The optical catalog of NGC 2516 was built from the list of 1254 photometric
members compiled by \citet{Jeffries01}. 
The member selection is based on the photometry in the B, V and I bands,
and 551 stars of this catalog lie in the combined field of view of the 6 XMM
observations. We have added to this catalog 42 cluster stars from 
\citet{DachsKab89} brighter than 
$V=9.7$, not present in the \citet{Jeffries01} sample.
This sample covers a $\sim 58\arcmin\times54\arcmin$ sky region, 
whereas the combined EPIC image covers only about $18\arcmin\times25\arcmin$,
centered in the core of the cluster (cf. Fig. \ref{img}). 
The number of cluster stars in the surveyed area is 593. In the text 
we will refer to the stars of this catalog as JTH and a running number, while 
for the optically bright stars we will use also DK acronym from 
\citet{DachsKab89}.

\begin{figure}
\begin{center}
\includegraphics[height=\columnwidth,angle=-90]{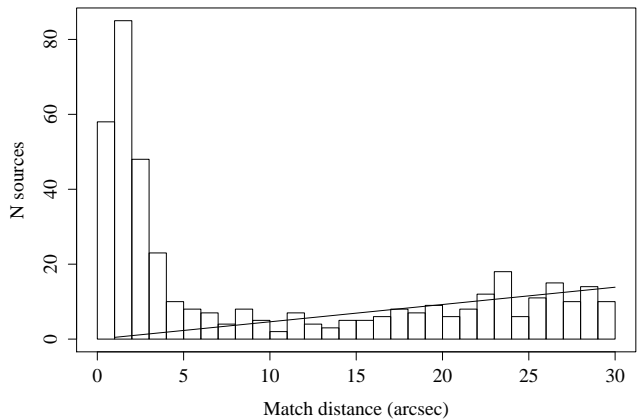}
\caption{\label{match} Histogram of offsets between X-ray sources
and the optical members positions of NGC 2516. The line traces 
the distribution of expected spurious matches.}
\end{center}
\end{figure}

We have matched the positions of X-ray sources with the optical 
coordinates of the composite catalog. No significant systematic offsets were 
found between optical and X-ray astrometry. 
By studying the distribution of the distances
of matched sources, we have chosen a maximun identification radius of 7$\arcsec$ 
where the number of observed matches is similar to the number of expected chance 
identifications.
Indeed we can estimate the expected number of chance matches
uncorrelated with true cluster members at a given threshold distance 
by taking into account the star density of the optical 
catalog in the EPIC field of view and the number of X-ray detections to 
be matched; 
it yields 12 fake matches at the chosen 7$\arcsec$ radius,  
by integrating the line traced in Fig. \ref{match}.
The value of adopted radius does not change significantly when matching the
X-ray sources at small and large offaxis positions, respectively. 
A list of 239 matches with the optical catalog was obtained; 
five X-ray sources were doubly matched with close pairs
in the optical catalog;
354 cluster members in the surveyed area remain undetected in X-rays 
and thus 
we have calculated upper limits to their X-ray count rates, fluxes and 
luminosities at the chosen detection threshold at the positions of undetected 
stars.
Tables \ref{xflu} and \ref{xupl} list the detected stars and the
upper limits for the cluster stars, respectively. 

We have estimated a conversion factor (CF) in the 0.3-7.9 keV band
between count rates (in units of MOS 1 instrument, as obtained by
the detection code) and unabsorbed fluxes,
derived by the spectral analysis of the bright X-ray sources of 
the cluster (see Sect. \ref{anspec}); the resulting CF, used for all the 
cluster X-ray sources, was $9.5\cdot 10^{-12}$ erg cnt$^{-1}$ cm$^{-2}$.
The 3 sigma uncertainty on CF  takes into
account the dependence on the spectrum hardness and amount to a $\pm 14\%$; 
it would shift systematically fluxes and luminosities by +0.06 and -0.07 dex, 
respectively.

To obtain the luminosities we assumed a distance to the cluster of 387 pc, as
adopted already by \citet{Jef97} and \citet{Damiani2003}. 
This distance is slightly larger than the {\em Hipparcos} satellite 
estimate (346 pc); by assuming this latter the luminosities would decrease 
by -0.05 dex. 
Table \ref{spdet} reports the number and the rate of detections for each 
spectral type (indeed for each range of B--V or V--I color as reported in 
the header of the Table), the number of stars in the field of view,
the medians of X-ray luminosities, also after correction for non members
contamination (see Sect. 3.3) and the minimum of detected luminosity.
The weakest source of the cluster is a M-type star, with a X-ray flux equal 
to $1.2\cdot 10^{-15}$  erg s$^{-1}$ cm$^{-2}$ and luminosity of $2.2 \cdot 10^{28}$ 
erg s$^{-1}$.
For comparison the Chandra survey of \citet{Damiani2003} led to a minimum
detected flux among cluster sources of $2.75 \cdot 10^{-15}$ erg s$^{-1}$ 
cm$^{-2}$ in a very similar band. 
The choice of the CF and the distance does not result in a systematic 
difference between the luminosities obtained in the present survey and 
that of \citet{Damiani2003} conducted with Chandra (cf. Fig. \ref{chandra}); in  
Sect. 4 we discuss in detail the comparison of luminosities estimated with 
Chandra and XMM-Newton.
\subsection{\label{err} Uncertainties on X-ray fluxes and luminosities}
We summarize here all the statistical and systematic errors 
that could affect our estimates of fluxes and luminosities. 
Count rates have relative errors in 5\%--30\% range
(as reported in Table \ref{xdet}).
The CF has a 14\% uncertainty at 3$\sigma$ level from fit of 
fluxes vs. count rates, essentially due to the effect of source spectrum 
temperature on CF. 
The scaling of pn to MOS introduces an uncertainty of $\sim$11\% i.e. 
fluxes and luminosities can be systematically higher or lower by -0.03 
or +0.015 dex, respectively.
The different distance estimates given in literature should systematically 
shift the X-ray luminosities by -0.05 (346 pc, Hipparcos distance) or +0.022 
(for a distance of 407 pc, \citealp{Terndrup02}). 
All these sources of uncertainties could balance themselves,
hence we guess that the errors in Table \ref{xflu} could suffer of a 
systematic additional uncertainty of $\sim \pm$ 0.07 dex in fluxes and 
$\pm$0.08 dex in luminosities.

\begin{table*}
\begin{center}
\caption{\label{spdet} Number of X-ray detections, number of stars in EPIC field of
view,  medians of $\log \mathrm{L}_\mathrm{X}$ distributions (we report also the values 
corrected for non member contaminants in G-, K- and M-type stars, see Sect. 3.3)
and minimum detected luminosity
for  B, A-, F-, G-, K- and M-type stars of NGC 2516. 
The selection of the spectral types is reported in the Table header. 
In the cases where the rate of detection is $< 50\%$ only an upper limit
to the median is reported.}
\vspace{0.1cm}
\resizebox{\textwidth}{!}{
\begin{tabular}{l l l l l l l}\hline \hline
Sp. Type & B & A & F & G & K & M \\ 
   &  $(B-V)_0 < 0$ & $0 \leq (B-V)_0 \leq 0.3$ & $0.3 \leq (B-V)_0 \leq 0.5$ & $0.5 \leq (B-V)_0 \leq 0.8^\mathrm a $ & $0.93 \leq (V-I)_0 \leq 2.2$ &  $2.2 \leq (V-I)_0 \leq 5.0$ \\ \hline
N$_\mathrm{det}$/N$_\mathrm{tot}$ & 13/38 & 16/49 & 24/37 & 52/89 & 74/182 & 59/194 \\
Detect. Rate & 34\% & 33\% & 65\% & 58\% & 41\% & 30\% \\ 
$\log \mathrm L _{\mathrm X}$ median & $\leq 28.62$ & $\leq 28.44$ & 29.19 & 28.95 & $\leq$ 28.70 & $\leq$ 28.62 \\
$\log \mathrm L _{\mathrm X}$ med.$_\mathrm{no cont.}$ & ... & ... & ... & 29.04 & $\leq$ 28.82 & $\leq$ 28.68 \\
Min Det $\log \mathrm L _{\mathrm X}$ & 28.85 & 28.60 & 28.71 & 28.55 & 28.5 & 28.34 \\ \hline
\end{tabular}
}
\end{center}
$^a$ Combined with: $(V-I)_0 \leq 0.93$
\end{table*}

\subsection{\label{anspec} Spectral analysis}
We used EPIC data to investigate the spectral 
properties of the NGC 2516 stellar coronae in terms of relevant temperatures and
emission measures. 
The pn has the highest efficiency in collecting X-ray photons among the 
three EPIC instruments, hence we have used the data from each pn exposure
by fitting simultaneously the X-ray spectra observed in different exposures. 
\footnote{In some cases we have excluded 
spectra with very few photons so for each source we do not fit simultaneously 
always all the spectra from the six observations.}
In  order to make a spectral analysis with global fitting technique. 
we have accumulated the spectra for source with more than 500 total counts 
in the sum of datasets.
For 31 sources with more than 1000 counts in the summed image, the photons were 
extracted within a 30$\arcsec$ radius circular region centered at the position 
of the source, and the background spectrum was accumulated in annuli (with 
radii of 35$\arcsec$ and 45$\arcsec$, respectively). 

For 7 sources with counts between 500 and 1000 total 
counts we chosen to accumulate in a 20$\arcsec$ radius region and to fit 
spectra without background subtraction, having considered that background 
photons are 
reduced to 44\% in this area with respect to the area of radius 30$\arcsec$, 
whereas the source encircled energy is reduced from 80\% to $\simeq 70\%$.
In this way we expect to have reduced the contamination of 
background spectrum  and, at the same time, to save most of the source photons 
without increasing uncertainties due to background subtraction.
We have tested this procedure in a few cases finding that the background 
subtraction increases by more than 30\% the relative error of fitted parameters,
while the best fit temperature slightly decreases within the uncertainty range.
The background photons are $\sim60\pm30$ in the 20$\arcsec$ extraction area,
the actual number depending on the source off-axis distance.
We have included also sources with only 500 counts to reduce possible biases
introduced by considering only higher count statistics hence higher 
activity stars.

The spectra of sources with highest statistics were fit in most cases with a 
sum of two APEC models plus photoelectric absorption. After having tried
models with free parameters the absorption and the abundance, 
we have decided to fix the column density value to 
N$_H = 8\cdot 10^{20}$ cm$^{-2}$, coherently with estimates from photometry 
\citep{Jeffries01}, and the abundances at 0.3 Z$_{\sun}$ because we noticed 
that under-solar abundances are always obtained, leaving thus as
free parameters the two temperatures and emission measures. 
The spectra of low statistics sources were fit with one APEC model, with 
free parameters the temperature, abundance and emission measure and a  
fixed absorption at N$_H = 8\cdot 10^{20}$ cm$^{-2}$, as chosen before.
Results from spectral fitting are reported in Table \ref{specres} and discussed
in the Sect. 3.2; errors refer to the 10\% -- 90\% $\chi^2$ confidence range.

\begin{table*}
\renewcommand{\arraystretch}{1.0}
\caption{\label{specres} Temperatures and Emission measures estimated from 
 global fit of coronal spectra. Models with 1-T or 2-T components have been used,
 errors refer to 10\%--90\% $\chi^2$ confidence range.}
\begin{tabular}{l l l l l l l}\hline \hline
Num & B--V & T$_1$ (keV) & T$_2$ (keV) & $\log$ EM$_1$ (cm$^{-3}$) & $\log$ EM$_2$ (cm$^{-3}$)  & $\chi^2$/dof \\ \hline
9465   &   0.818   &   0.72$^{0.07}_{-0.08}$   &   1.69$^{0.5}_{-0.3}$   &   52.93$^{0.04}_{-0.17}$   &   53.06$^{0.02}_{-0.17}$ & 2.7/88\\
9676   &   0.903   &   0.34$^{0.09}_{-0.07}$   &   1.14$^{0.13}_{-0.11}$   &   52.79$^{0.02}_{-0.2}$   &   52.94$^{0}_{-0.13}$ & 2.4/95\\
7678   &   0.928   &   0.37$^{0.13}_{-0.08}$   &   1.59$^{0.7}_{-0.3}$   &   52.6$^{0.04}_{-0.24}$   &   52.63$^{0.03}_{-0.18}$& 2.3/25\\
7782   &   1.417   &   0.28$^{0.1}_{-0.09}$   &   1.25$^{0.5}_{-0.3}$   &   52.69$^{0.05}_{-0.5}$   &   52.99$^{0.04}_{-0.13}$ & 4.5/7\\
9175   &   0.585   &   0.66$^{0.09}_{-0.07}$   &   --   &   53.2$^{0.1}_{-0.2}$   &   -- & 3.5 / 15\\
8893   &   1.433   &   0.32$^{0.3}_{-0.06}$   &   1.29$^{0.15}_{-0.15}$   &   52.51$^{0.08}_{-0.2}$   &   52.73$^{0.03}_{-0.1}$ & 2.3 / 42\\
11233   &   0.834   &   0.6$^{0.08}_{-0.2}$   &   1.2$^{0.2}_{-0.2}$   &   52.91$^{0.09}_{-0.14}$   &   52.86$^{0.06}_{-0.16}$ & 1.3 / 55\\
6649   &   0.817   &   0.56$^{0.24}_{-0.25}$   &   1.3$^{0.3}_{-0.2}$   &   52.22$^{0.4}_{-0.2}$   &   52.78$^{0.08}_{-0.2}$  & 1.1 / 23\\
7585   &   0.719   &   0.6$^{0.07}_{-0.2}$   &   2.04$^{1.1}_{-0.6}$   &   52.68$^{0.07}_{-0.3}$   &   52.68$^{0.04}_{-0.13}$ & 1.4 / 26\\
4598   &   0.861   &   0.7$^{0.1}_{-0.1}$   &   2.1$^{0.9}_{-0.5}$   &   52.96$^{0.04}_{-0.2}$   &   53.16$^{0.1}_{-0.21}$ & 1.3 / 28\\
9140   &   0.736   &   0.74$^{0.07}_{-0.14}$   &   1.7$^{2.5}_{-0.7}$   &   52.85$^{0.04}_{-0.2}$   &   52.54$^{0.2}_{-0.6}$ & 1.35 / 26\\
7864   &   0.603   &   0.7$^{0.4}_{-0.4}$   &   1.6$^{8.4}_{-0.5}$   &   52.5$^{0.2}_{-0.7}$   &   52.8$^{0.09}_{-0.5}$ & 0.3 / 6\\
8997   &   0.981   &   0.8$^{0.2}_{-0.1}$   &   --   &   52.85$^{-0.01}_{-0.11}$   &   -- & 2.0 / 38 \\
8458   &   0.533   &   0.54$^{0.04}_{-0.06}$   &   --   &   52.92$^{-0.01}_{-0.1}$   &   -- & 1.2 / 47 \\
8099   &   0.56   &   0.61$^{0.05}_{-0.05}$   &   --   &   52.81$^{-0.02}_{-0.11}$   &   -- & 2.1 / 20\\
9054   &   0.691   &   0.32$^{0.06}_{-0.09}$   &   0.9$^{0.09}_{-0.15}$   &   52.73$^{0.1}_{-0.1}$   &   52.65$^{0.09}_{-0.06}$ & 1.6 / 49\\
7667   &   0.195   &   0.60$^{0.4}_{-0.04}$   &   --   &   52.66$^{0.05}_{-0.05}$   &   -- & 1.7 / 26\\
10046   &   0.99   &   0.31$^{0.05}_{-0.05}$   &   1.5$^{0.3}_{-0.3}$   &   52.81$^{-0.01}_{-0.3}$   &   52.93$^{0.02}_{-0.1}$ & 0.7 / 16\\
8529   &   0.583   &   0.64$^{0.06}_{-0.04}$   &   --   &   52.85$^{-0.03}_{-0.1}$   &   -- & 2.1 / 24\\
8886   &   0.113   &   0.32$^{0.17}_{-0.2}$   &   3.5$^{1.7}_{-1.0}$   &   52.3$^{1.3}_{-0.7}$   &   52.9$^{0.1}_{-0.1}$ & 0.7 / 22\\
6029   &   0.674   &   0.4$^{0.2}_{-0.1}$   &   1.3$^{1.1}_{-0.3}$   &   52.92$^{0.05}_{-0.2}$   &   52.7$^{0.01}_{-0.3}$& 1.8 / 19\\
8634   &   0.943   &   0.38$^{0.14}_{-0.07}$   &   1.21$^{0.14}_{-0.15}$   &   52.79$^{0.01}_{-0.2}$   &   52.92$^{-0.02}_{-0.16}$& 1.0/34\\
9835   &   0.988   &   0.14$^{0.11}_{-0.13}$   &   0.94$^{0.13}_{-0.12}$   &   52.9$^{1.2}_{-0.6}$   &   52.91$^{-0.03}_{-0.2}$& 2.4/14\\
10817   &   0.525   &   0.59$^{0.16}_{-0.17}$   &   11$^{6}_{-6}$   &   53.02$^{0.05}_{-0.3}$   &   53.48$^{0.05}_{-0.07}$& 1.4/16\\
8641   &   1.063   &   0.37$^{0.13}_{-0.07}$   &   0.99$^{0.09}_{-0.09}$   &   52.91$^{0.02}_{-0.23}$   &   52.99$^{0.01}_{-0.12}$& 1.4 /64\\
7743   &   0.655   &   0.73$^{0.07}_{-0.09}$   &   --   &   52.82$^{-0.01}_{-0.13}$   &   -- & 1.0 / 12\\
DK23   &   0.06   &   0.56$^{0.06}_{-0.12}$   &   --   &   51.6$^{1.0}_{-0.2}$   &   -- & 1.6 / 21 \\
DK44B   &   -0.02   &   0.73$^{0.07}_{-0.11}$   &   --   &   52.68$^{0.05}_{-0.05}$   &   -- & 1.5 / 24\\
DK56   &   -0.09   &   0.8$^{0.2}_{-0.1}$   &   3$^{0.7}_{-0.4}$   &   52.56$^{0.09}_{-0.2}$   &   53.29$^{0.04}_{-0.04}$ & 0.9 / 99 \\
DK73   &   0.02   &   0.6$^{0.1}_{-0.07}$   &   --   &   52.69$^{0.07}_{-0.05}$   &   -- & 1.5 /19 \\
DK62   &   0.03   &   0.61$^{0.05}_{-0.2}$   &   1.4$^{0.5}_{-0.3}$   &   52.76$^{0.08}_{-0.18}$   &   52.81$^{0.11}_{-0.17}$ & 1.5 / 39\\
6689   &   0.329   &   0.73$^{0.12}_{-0.11}$   &   --   &   53.43$^{0.11}_{-0.16}$   &   -- & 1.4 / 10 \\
9852   &   0.659   &   0.54$^{0.06}_{-0.07}$   &   --   &   53.0$^{0.2}_{-0.5}$   &   -- & 2.2 / 19 \\
7650   &   0.62   &   0.6$^{0.1}_{-0.1}$   &   --   &   53.0$^{0.2}_{-0.3}$   &   -- & 1.1 / 6 \\
8660   &   0.571   &   0.59$^{0.08}_{-0.1}$   &   --   &   53.1$^{0.3}_{-0.2}$   &   -- & 3.8 / 12\\
10040   &   1.034   &   0.76$^{0.09}_{-0.1}$   &   --   &   53.22$^{0.15}_{-0.15}$   &   -- & 0.9 / 12\\
9882   &   0.185   &   0.65$^{0.06}_{-0.06}$   &   --   &   53.1$^{0.2}_{-0.3}$   &   -- & 1.1 / 13 \\
8967   &   0.996   &   0.75$^{0.17}_{-0.17}$   &   --   &   53.24$^{0.2}_{-0.17}$   &   --& 0.7 / 5\\
\hline
\end{tabular}
\end{table*}

\section{Discussion}
\subsection{X-ray emission along the Main Sequence}
\begin{figure*}
\centering
\includegraphics[height=\textwidth,angle=-90]{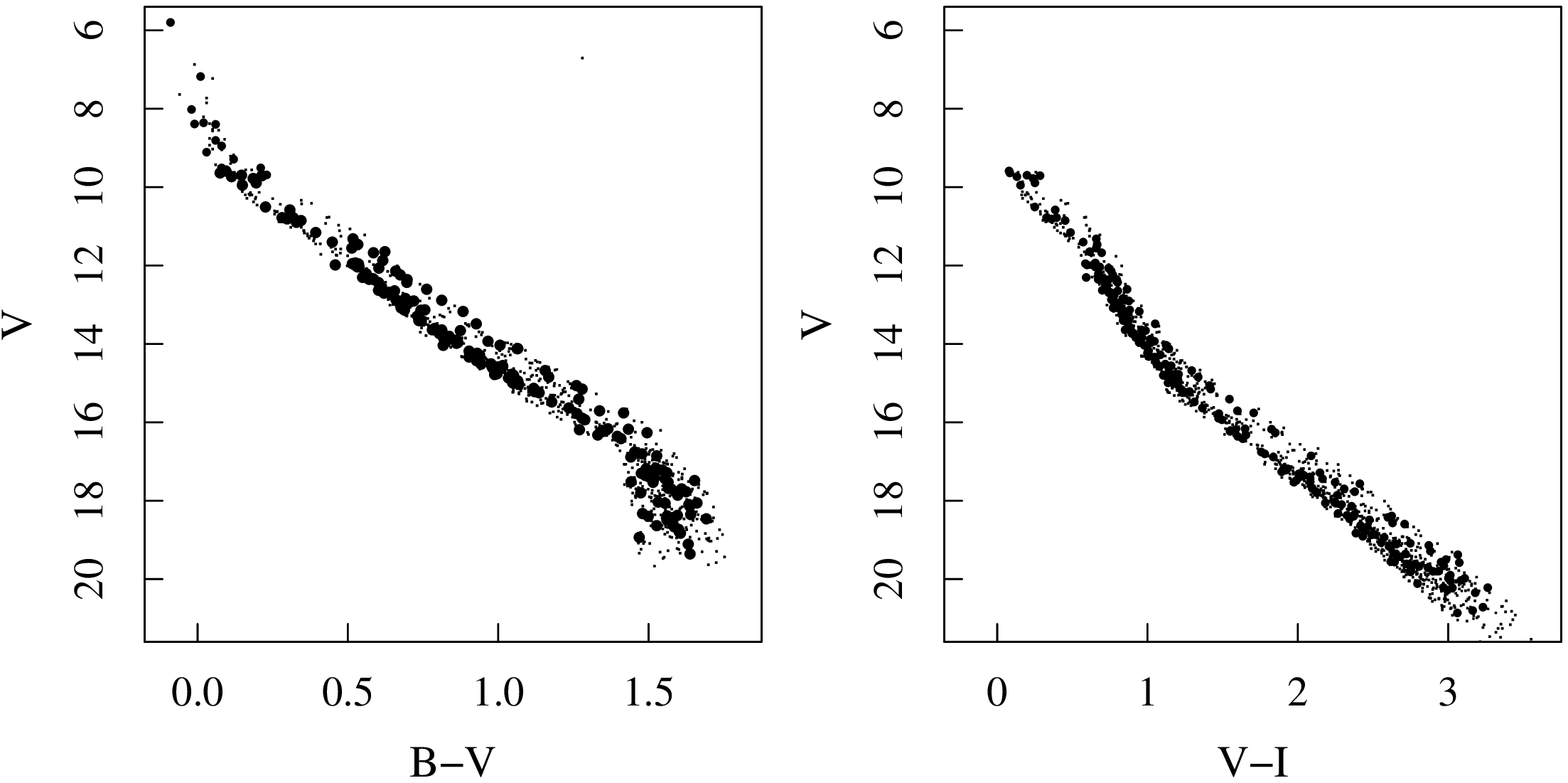}
\caption{\label{vbv} Color-Magnitude diagrams in B, V, I bands of NGC 2516 stars
(small dots) from \protect{\citet{Jeffries01}} and \protect{\citet{DachsKab89}}. 
Big dots are X-ray detected stars.}
\end{figure*}

Fig. \ref{vbv} shows the Color-Magnitude Diagrams (CMD) in B, V, I bands 
for the photometric cluster members. 
Big dots are X-ray detections, small dots are the undetected cluster 
stars. Hotter stars are shown only in the left panel (CMD with B--V) while 
cooler stars with unreliable B--V are shown only in the CMD with V--I color index.
The X-ray detections span the Main Sequence of the cluster 
from B type stars down to cool, low mass M-type stars.

\begin{figure}
\includegraphics[height=\columnwidth,angle=-90]{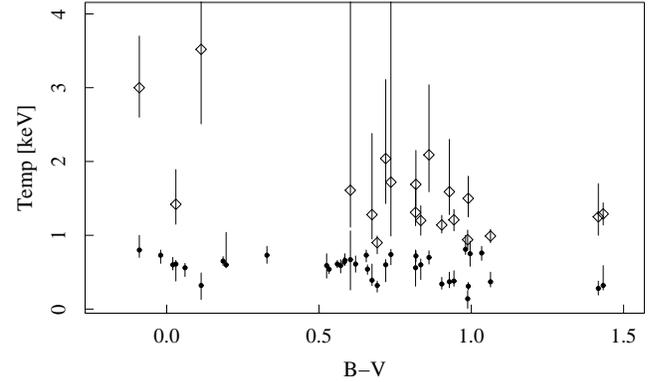}
\caption{\label{tem} Coronal temperatures as a function of B--V color.
Filled dots refer to the cool component whereas empty diamonds refer to the hot 
component.}
\end{figure}

\subsection{X-ray spectral properties}
The sample of stars of which we have analyzed the spectra are constituted 
by solar G- and K- type stars and by hot, massive B- and early A- type stars. 
We were unable to analyze the spectra of stars with B--V$\sim0.4$ 
(late-A spectral type stars) because of their lower X-ray 
fluxes  and hence poor statistic spectra. Although
the models we used are too simple and unrealistic, 
a distribution of plasma temperatures being more realistic,
we may determine the dominant temperatures of the coronae as usually done 
with low resolution X-ray spectra.

In Fig. \ref{tem} we plot the temperatures derived from global fitting as a 
function of B--V color. 
We were able to fit most of the spectra of solar type stars with 2 temperature 
models. High energy tails of spectra are very scarce of photons, consequently
hot components have been estimated with large errors.
The temperatures are in 0.3--0.7 keV and 1.0--2.0 keV range, 
respectively, and are similar to those detected in the coronae of Pleiades 
and Blanco 1 open clusters, which are nearly coeval to NGC 2516 
\citep{Briggs2003,Pilli2004}. The ratio of hot to cool emission measure in 
2-T models suggests that the two thermal components are comparable,
the ratio being between 0.5 and 3.0 with a median of 1.3.

Spectra of all but 3 early type stars (B--V $\la 0.5$) have low count 
statistics: we may  obtain a good fit of them with 1-T models. 
In this case the temperatures range around 0.6 keV, 
suggesting that the spectra are softer than those of solar mass stars.

The star DK 56 (a binary system with a B2V star
and an unresolved late type companion) was bright enough in each 
observation to fit its spectrum with a 2-T model. We found a hot 
spectrum with temperatures of 0.8 and 3.0 keV and a spectrum shape 
similar to those of solar type stars. In the case of a wind driven X-ray 
emission, typical of massive stars, the spectrum is expected to be softer
than in the case of coronae of solar mass stars. Therefore, based on the 
spectrum shape and temperatures, we suggest that the X-ray emission of 
DK 56 is likely 
due or strongly affected by the late type companion.
This is a conclusion opposite to that of \citet{Jef97} who, on the
basis of its high X-ray luminosity, attributed all the X-ray emission of DK56 
to the massive primary. 
This system is a blue straggler of the cluster,
younger than $\sim 25$Myr \citep{DachsKab89}. Based on this
age estimate, these authors discussed the hypothesis of subsequent episodes
of star formation in NGC 2516 determined by the crossing of the cluster through
the Galactic Plane or the Gum Nebula.
Given this age estimate, the high X-ray luminosity of order of 10$^{30}$ 
erg s$^{-1}$ is not an atypical value for solar-type stars of age of 25 Myr 
or less.
\begin{figure*}
\includegraphics[height=\textwidth]{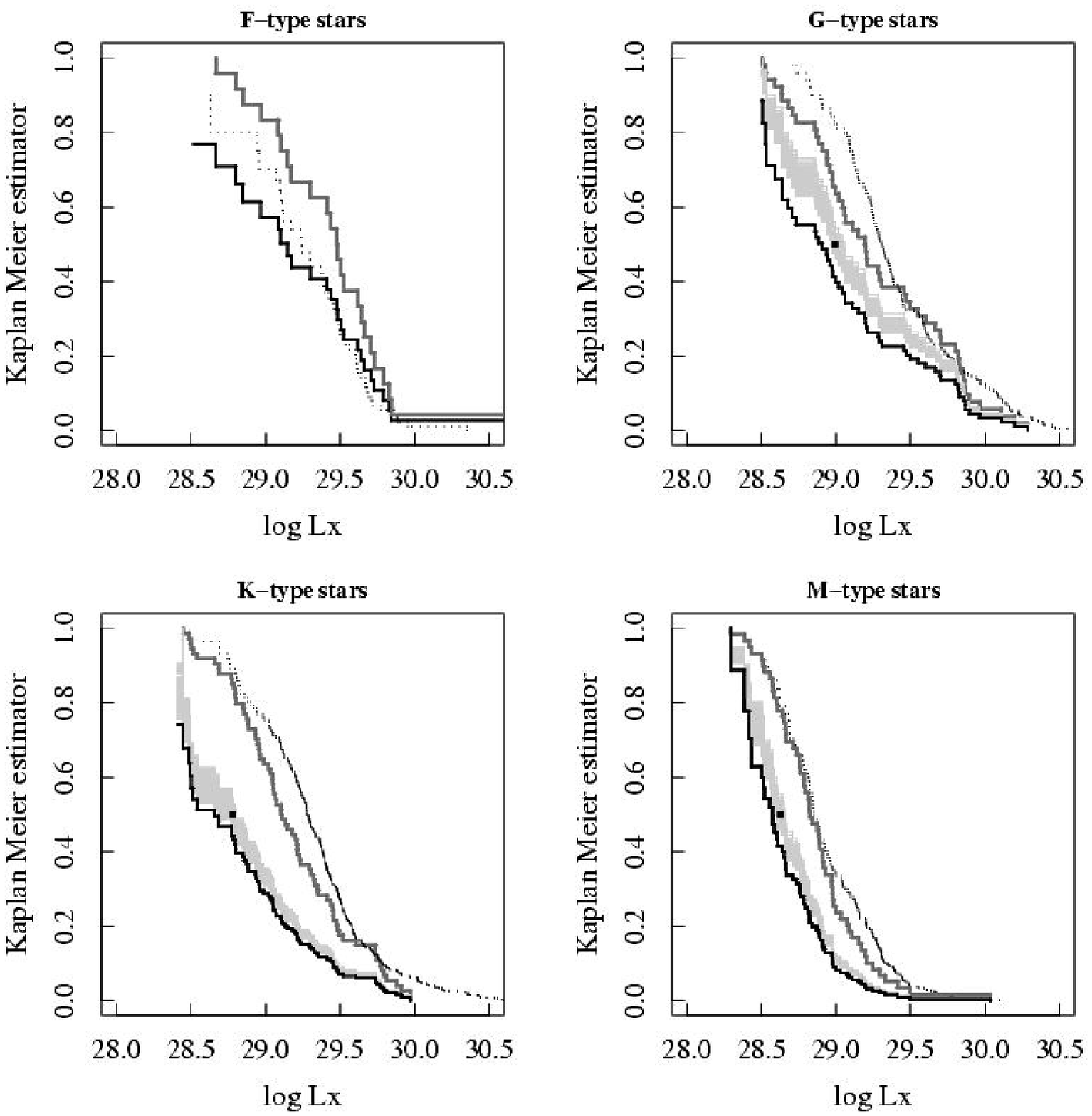}
\caption{\label{xlf} Cumulative distribution functions of X-ray luminosities
of NGC 2516 stars (solid black line) with different spectral types, 
evaluated with Kaplan Meyer estimators. 
Dotted lines are the distributions of Pleiades \protect{\citep{Giusi999}}. 
Solid gray line in G-, K- and M-type stars are the distribution of 
detections alone.
Gray area in G-, K- and M-type panels marks the envelop of 
simulated Kaplan-Meier estimators after statistical correction for contaminants;
the black square inside each area is the average of their medians.
}
\end{figure*}
\subsection{\label{xlfs} The X-ray luminosities of NGC 2516 late type stars}
We have calculated the Kaplan-Meier estimators \citep{Feigelson85} 
of the distribution of X-ray luminosities for F, G, K and M spectral types.
The distributions take into account the 
presence of the upper limits in $\log \mathrm L_\mathrm X$ that censorize 
the sample. When the lowest values are upper limits the 
distribution does not reach unity. 
Fig. \ref{xlf} shows the X-ray distribution functions for F-, G-, K-
and M-type stars of NGC 2516 (solid lines); for comparison we have added the 
ROSAT data of Pleiades (\citealp{Giusi999}, dotted line) and the distribution
considering only detections of NGC 2516 (solid gray line). 
This is an upper bound to the true distribution which must include undetected
members.
The Kaplan-Meier estimators of Pleiades and NGC 2516 are different
\footnote{By applying several two sample tests, the difference between 
the two curves is significant at a level $\ga$ 99.9\%.}. 
In particular, the luminosities of NGC 2516 of
G-, K- and M- type stars are systematically lower than those of the Pleiades, 
the difference being larger for G- and K-type stars where the
contamination of the optical catalog is higher than in the other spectral
ranges,
as stated by \citet{Jeffries01}. Especially among K-type stars the 
Kaplan-Meier estimator and cumulative distribution of X-ray
detections alone show the largest difference.

The contamination by less active field stars among G- and K-type stars in
NGC 2516, hampers the evaluation of L$_\mathrm X$ distributions of NGC 2516. 
However, also the X-ray luminosities of detections alone
of G-, K- and M-type stars are systematically lower than the luminosities
of the Pleiades at a level $\geq 99\%$ as it results by applying several 
two sample tests.
The distribution of detections alone shown in Fig.
\ref{xlf} may be interpreted as an upper limit for the {\em true} 
cumulative distribution of X-ray luminosity. 
If we were able to take into account the upper limits of only 
{\em true} cluster members, the Kaplan-Meier estimators should be at
lower luminosities with respect to the distribution of detections alone.

We have evaluated how many field stars are expected to fall among our 
X-ray detections. We focused on the main sequence field stars expected
to be in a volume determined by the photometric selection of cluster 
Main Sequence and the sky area of our survey.
By using the stellar density given in \citet{Allen2000}, Table 19.14 
(valid for stellar density on the galactic plane, NGC 2516 is $\sim$100 
pc above the Galactic Plane) we estimated 
11 and 19 G and K-type main sequence field stars 
comprised in that volume and not related to the cluster. 

From the X-ray luminosity distributions reported in \citet{Schmitt95} and 
\citet{Schmitt97} G-type field stars have 
$\log \mathrm L_\mathrm X/(\mathrm{erg}\ \mathrm{s}^{-1})$ 
in the range 26.5--29.5 with median 27.3; K-type stars occupy a narrower range 
(27.1--28.4) with a median of $\log \mathrm L_\mathrm X/(\mathrm{erg}\ 
\mathrm{s}^{-1})=$ 27.6. 
M-type stars in the solar neighborhood have luminosities $25.5 \leq 
\log \mathrm L_\mathrm X/(\mathrm{erg}\ \mathrm{s}^{-1}) \leq 29.1$ 
and a median of 
$\log \mathrm L_\mathrm X/(\mathrm{erg}\ \mathrm{s}^{-1}) \sim 27$.
Hence we estimated that less than 3 field stars in each spectral type should 
have been detected among the photometric sample 
(at the same time a sample of 10 to 18 stars should be undetected 
in the surveyed volume).
More distant giants should not be detected and thus they should not affect 
the distribution of detections alone.
The expected field star detections are thus very few compared to the number of 
detected cluster members in each sample (52 G-, 74 K- and 59 M-type 
detections), thus the distribution of detections alone should not be affected 
by field stars. 

We have taken into account the expected fraction of contaminants 
given by \citet{Jeffries01} in order to estimate {\em unbiased} X-ray luminosity
distribution functions of NGC 2516 G-, K- and M-type stars.
By following the method used in \citet{Damiani2003}, we excluded a fraction of upper limits
from the X-ray luminosity sample in each spectral type according to the 
membership probability and assuming that the contaminants yield only upper 
limits; then we calculated the Kaplan-Meier estimator. 
This simulation procedure was iterated 500 times producing thus the bunch of 
curves plotted in light gray in Fig. \ref{xlf} and which define the region 
where more likely the {\em true} distribution function should lie.
Only in the case of G-type stars we observe a larger difference between the 
curve with all upper limits and the family of contaminant corrected curves. 
In the K- and M-type stars the large number of upper limits is less affected 
by the statistical correction for contaminants.
We point out that in all cases the gray region is at lower X-ray luminosities 
with respect to the Pleiades cumulative distributions.

\begin{figure}
\includegraphics[height=\columnwidth,angle=-90]{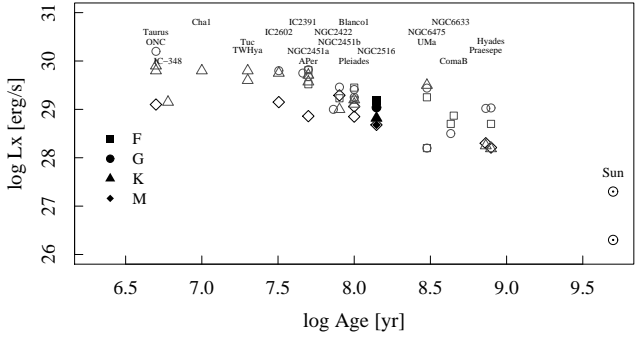}
\caption{\label{lxage} $\log \mathrm{L}_\mathrm X$ vs. age for different open 
clusters, star forming regions and the Sun. Open symbols: data of all but 
NGC 2516 clusters from \protect{\citet{Guedel04}}; solid symbols:
medians of NGC 2516 F-, G-, K- and M-type stars (contaminants corrected, 
see. Sect. 3.3) from the present work.}  
\end{figure}

The averages of the 0.5 quantile
of each simulated curve are marked in each panel with a black square; 
the $\log$ L$_ \mathrm X /(\mathrm{erg}\ \mathrm{s}^{-1})$ values are 
29.04, 28.82 and 28.68 for G-, K- and M-type stars, respectively;
the width of the gray region at the 0.5 quantile is $\sim$0.05 dex which 
we assume as the uncertainty on the previous averages.
\citet{Guedel04} and \citet{Favatarev03} discuss X-ray emission as a function
of age for stars from young stellar objects to aged open clusters and the Sun.
We have plotted in Fig. \ref{lxage} the medians of $\log \mathrm L_\mathrm X$
for F-, G-, K- and M-type stars reported by \citet{Guedel04} and 
our $\log \mathrm L_\mathrm X$ medians (after contaminant correction) for the same
spectral types as a function of age. We have assumed here an age of 140 Myr for NGC 2516.
It is evident the decrease of the mean level of X-ray luminosities with
age, after a plateau at $\sim10^{30}$ erg s$^{-1}$ which extends from Orion 
and Taurus Star forming Region ages (few Myr) to Alpha Persei 
cluster age (50 Myr). 
The X-ray under-luminosity of NGC 2516 with respect to the nearly coeval
Pleiades remains evident, also taking into account the statistical correction 
for non member contamination and any systematic uncertainty discussed in Sect. 2.2.

The causes of the different level of luminosity of solar type stars in NGC 
2516 and the Pleiades could be attributed to several factors. 
The rotation is a key parameter for
X-ray emission \citep{Pallavicini81} up to rotation as fast as v 
sin$i\sim 20$ km/s,
when the L$_\mathrm X$ -- rotation relation flattens with respect to rotation. 
Furthermore, angular momentum losses due to magnetic braking link stellar
rotation to age and X-ray activity. 
A different distribution of rotation rates in the Pleiades and NGC 2516 could 
explain the low X-ray luminosity observed in the latter cluster.
\citet{Jef98} studied the stellar rotation of solar type stars of 
NGC 2516, finding a lower rotational rate in F- early and G- type stars with 
respect to the Pleiades whereas no clear difference between these two clusters
was found among K-type stars. 
However \citet{Terndrup02} concluded that no significant 
difference in rotation distribution is present between the Pleiades and NGC 
2516, if this latter is assumed slight older than Pleiades.
The under-luminosity of NGC 2516 in X-rays seems arising essentialy from
the slight difference of age between the two clusters 
(80--100 Myr for Pleiades, 140 Myr for NGC 2516),
although the past history of NGC 2516, 
its location and kinematics could have played a role in determining 
peculiar forming conditions, perhaps different from those of the Pleiades with
an influence on the X-ray emission evolution.

\subsection{L$_\mathrm X$ to L$_\mathrm{bol}$ ratio}
An indicator of activity is the ratio of X-ray to stellar 
bolometric luminosity, \lxlbol. Because of the dependence of X-ray 
luminosity on rotation, the \lxlbol\ ratio follows a power law relation 
with index -2 with rotational period and the Rossby number, i.e. 
the ratio between rotational period and convective turnover time
in {\em non saturated} stars (see \citealp{Patten96}, \citealp{Randich00},
\citealp{Nicola2003}).
ROSAT observations of young open clusters in the band 0.1--2.4 keV 
have also shown that low mass stars exhibit a saturation of this ratio at the 
level of $\sim$ -3, i.e. the X-ray luminosity reaches at most 1/1000 of 
bolometric luminosity of the star. Furthermore, saturation seems to occur 
at earlier types in younger stars (B--V$ = 0.7$ or late G-type stars 
at the age of the Pleiades, B--V $ = 1$ or mid K-type stars at 220 Myr, 
\citealp{Prosser95}).

\begin{figure*}
\includegraphics[height=\textwidth,angle=-90]{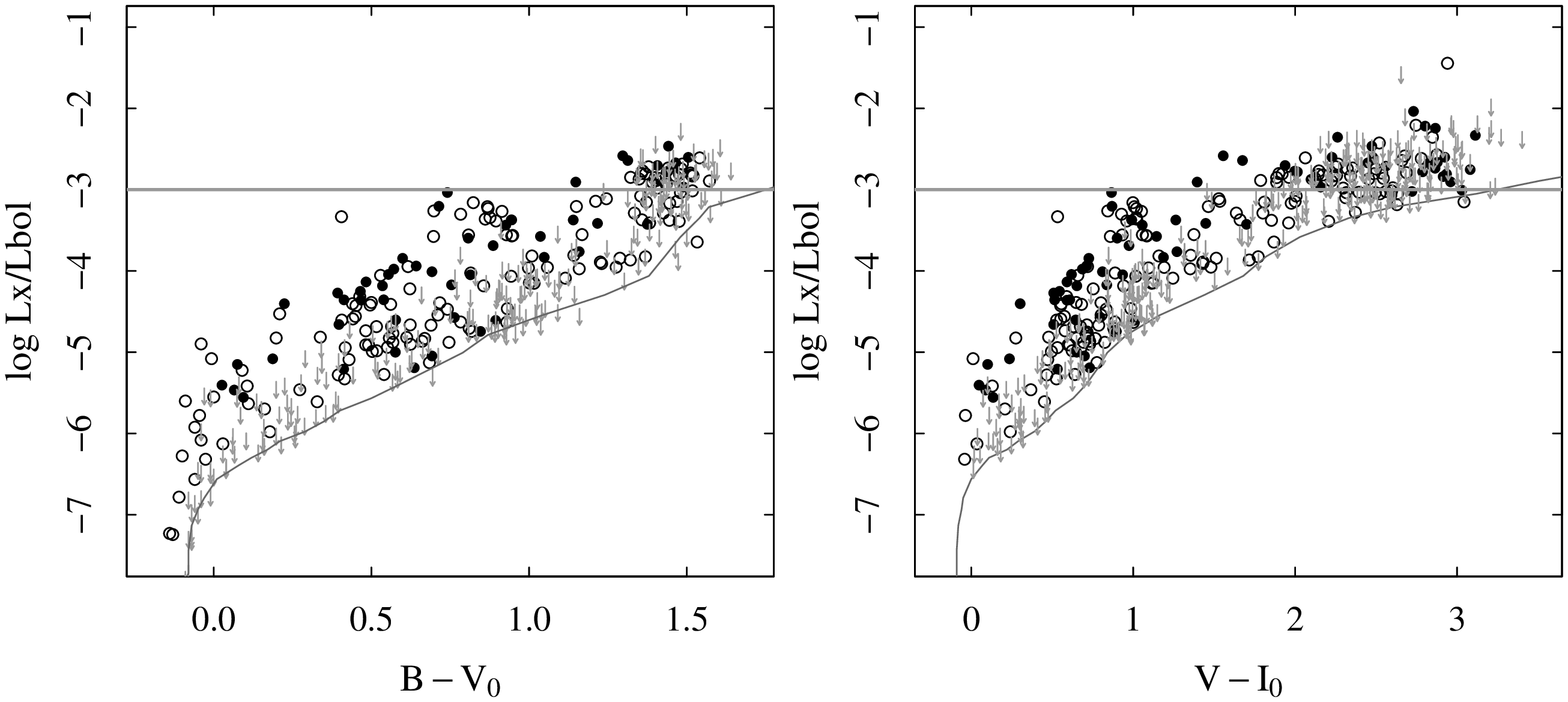}
\caption{\label{lxlb} Logarithmic X-ray to bolometric luminosity ratio 
of NGC 2516 stars. Open circles: detected single stars, filled circles: 
detected binaries, vertical arrows: upper limits}. Lines below data points
roughly trace the detection limit at the cluster distance 
($\log L_\mathrm X \sim 28.4$).
\end{figure*}

We calculated the bolometric luminosities of NGC 2516 by interpolating 
L$_\mathrm{bol}$ as a function of (B--V) based on a isochrone model with 
Z=0.02 and age of 140 Myr \citep{Siess2000};
for stars with (B--V)$_0\geq 0.3$ we used (V--I) color instead of (B--V).
Fig. \ref{lxlb} shows \lxlbol\ of NGC 2516 stars versus their B--V$_0$ and 
V--I$_0$ colors, with open and filled circles for single and binaries, 
respectively.
We did not correct L$_\mathrm{bol}$ for unresolved binary systems.
These unresolved binaries trace an upper sequence in the plots.
Arrows represent upper limits and the bottom curve traces approximately the
detection limit at the cluster distance ($\log$L$_\mathrm{X}\sim 28.4$). 
A spread of $\sim$1.4 dex is observed among detections in $0\leq B-V \leq1.5$ 
where \lxlbol\ may reach values up to -2.4. 
The highest point at $\sim$-1.4 is due to a very cool star 
(V--I= 3.06, id. 6160 in \citealp{Jeffries01} catalog): however this 
star is very close to the edge of field of  view and was observed only 
during the first observation, hence its X-ray luminosity could be quite 
uncertain. 

The \lxlbol\ ratio flattens at (V--I)$_0 \sim 1.5$ or (B--V)$_0 \sim 0.7$,
which corresponds to the spectral type K5, according to the 
color--temperature calibration by \citet{kenhart95} and we observe a saturation 
value of \lxlbol\ $\sim -2.5$
instead of -3. A comparison with L$_\mathrm{bol}$ used by \citet{Jef97} shows 
that our L$_\mathrm{bol}$ are on average lower by 0.1 dex for stars detected
in both surveys. 
This difference would produce essentially a shift toward high \lxlbol\ ratio, 
while the trend and the spectral type at which the saturation of \lxlbol\ 
occurs should not be changed. 

\section{Comparison with Chandra and ROSAT surveys}
\begin{figure}
\includegraphics[height=\columnwidth,angle=-90]{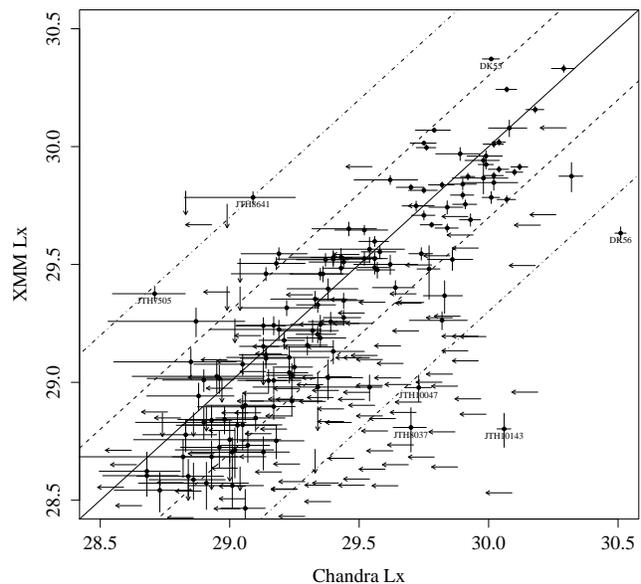}
\caption{\label{chandra} XMM-Newton vs. Chandra \protect{\citep{Damiani2003}}
measurements of X-ray luminosities of NGC 2516 stars detected at least once 
in the two surveys. Error bars are the statistical uncertainties;
the horizontal arrows 
mark  upper limits for undetected stars in Chandra, while vertical ones refer
to XMM-Newton undetected stars. Solid, dashed and dot-dashed lines trace
the equal, two and five times (and their reciprocal) ratios.
We indicate the names of a few stars with large variability, JTH are
from \citet{Jeffries01}, while DK refers to \citet{DachsKab89} }
\end{figure}

NGC 2516 has been observed several times with {\em Chandra} satellite 
between August 1999 and March 2001 with different instrument setups; 
a deep survey has been obtained by 
\citet{Damiani2003} combining together several exposures from ACIS and HRC
cameras. In that work 155 cluster stars have been detected in X-rays while
570 remained undetected. There are 535 stars falling both in the XMM-Newton
and Chandra fields of view; out of this sample we have detected 125 
stars with EPIC also detected with Chandra; further 100 stars are detected 
with EPIC whereas they were undetected in Chandra images. 
Only 17 stars are detected with 
Chandra but not with XMM-Newton; finally, 293 stars remain undetected in both 
surveys. In Fig. \ref{chandra} we show the comparison of Chandra and XMM-Newton
X-ray luminosities for the stars detected in at least one of these surveys. 
The plot shows that a scatter of a factor $\la$ 2 is observed in most 
cases; sometimes a difference of a factor 5 or more may be
present also in a few massive stars (DK 55 / HD 66167, B9.5V + A0V; 
HD 66194 / DK 56, sp. type B2IV).

The upper limits marked by arrows in the lower right corner of 
Fig. \ref{chandra} are stars undetected with Chandra but detected 
with XMM-Newton. 
This large number of upper limits is due to the sensitivity of our survey
being a factor $\sim$5 higher than the Chandra one.
It is worth to notice also that the maximum sensitivity
is comparable in both satellite instruments but in  Chandra the decrease
of sensitivity at large off-axis is more marked.
Both the large effective area of XMM-Newton telescopes and the more uniform 
point-spread-function (PSF) throughout the EPIC field of view result
in a higher efficiency of EPIC camera in detecting off-axis sources (thus in 
a larger sky area) with respect to ACIS and HRC detectors. 
Chandra has a higher spatial resolution near the center of the field 
of view, but its PSF profiles rapidly degrades at increasing 
off-axis thus reducing the efficiency in detecting faint sources.

\begin{figure}
\includegraphics[height=\columnwidth,angle=-90]{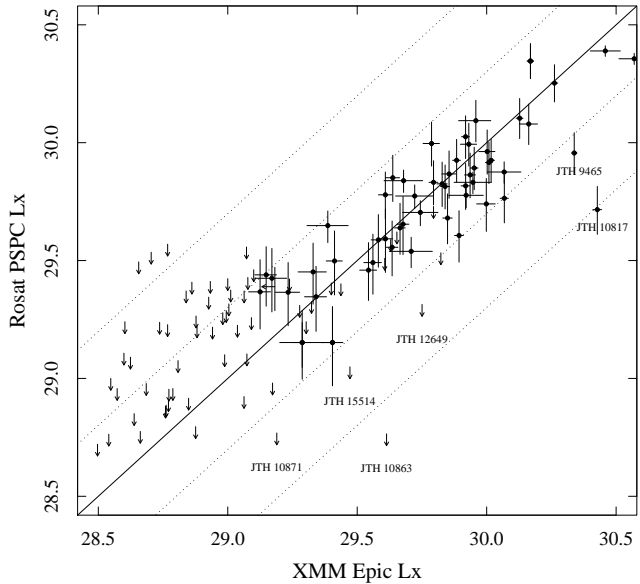}
\caption{\label{rosat} X-ray luminosities of XMM detected stars of NGC 2516
compared with values obtained with ROSAT satellite by \protect{\citet{Jef97}}. Arrows
mark upper limits due to ROSAT undetected stars. Continuos line and
dotted lines correspond to 1, 1/5, 1/2, 2 and 5 ratio between XMM-Newton and ROSAT 
luminosities, respectively.}
\end{figure}

The Sun activity is characterized by the 11-yr cycle.
The maximum of variation amplitudes in the Sun cycle amount to a factor 
greater than 20. 
In order to study the time variability on time scale of about 12 years
we have compared the X-ray luminosities of NGC 2516 stars falling in the field 
of view of ROSAT observations taken in 1993 and analyzed by \citet{Jef97}. 
This time scale should allow us to detect stellar cycles present in NGC 2516 
stars  with amplitudes and periods similar to those observed on the Sun.
Fig. \ref{rosat} shows the X-ray luminosities measured with
XMM-Newton and with ROSAT; XMM-Newton luminosities were recalculated 
coherently in the ROSAT band. Error bars for both XMM and ROSAT
 L$_\mathrm X$ are plotted, the latters obtained from \citet{Jef97}.
The lowest L$_\mathrm X$ detected with ROSAT is $\sim 10^{29}$ erg/s. 
Only two stars detected in both surveys, 
JTH 9465 (also JTX 138 in SIMBAD) and JTH 10817 (JTX 114),
changed the luminosity by a factor between 2 and 5, due to
flare variability during one of the six XMM observations. 
The variability of these two stars is discussed in \citet{Ramsay03}.
Other four stars (JTH 10863, 10871, 12649 and 15514), undetected in ROSAT,
are interesting because they are variable in XMM survey by more than a 
factor two with respect to their ROSAT upper limits.
Most of the stars have variations within a factor two.
The modest variability on long time scale has been reported also 
by \citet{Wolk04}, on a study of X-ray time variability of NGC 2516 based 
on Chandra data.
These variations are smaller than expected in the presence of
solar cycles and strongly suggest that cycles in young stars are absent 
or with different periods and/or small amplitudes than in the case of the Sun.

\section{New possible members and unidentified sources}
\begin{figure}
\includegraphics[width=\columnwidth,angle=-90]{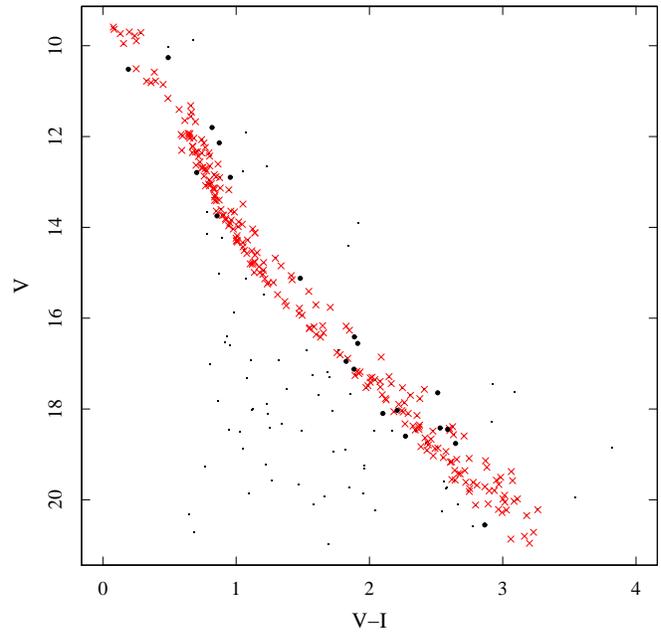}
\caption{\label{altri} Color magnitude diagram of X-ray sources matching the 
optical catalog of \protect{\citet{Jeffries01}}. Small dots: all sources, crosses: 
NGC 2516 stars, big dots: suggested new cluster members.}
\end{figure}

A number of X-ray detections have optical photometry consistent with the membership
of NGC 2516. On this basis these stars could be cluster members.
Fig. \ref{altri} shows the color magnitude diagram
of the X-ray sources from the Jeffries catalog.
Crosses are the members defined by \citet{Jeffries01}, 
big dots are a sample of stars that on the basis of X-ray observations 
could be new members of the cluster. 
By placing them at the cluster distance, their X-ray luminosities
range between $3.7\cdot\ 10^{28}$ and $1.5\cdot\ 10^{30}$ erg/s.
In Table \ref{lealtre} we list the main properties of these objects.

\begin{table*}
\renewcommand{\arraystretch}{1.1}
\caption{\label{lealtre} Properties of new suggested cluster members.
Columns are: id. number in \protect{\citet{Jeffries01}} catalog, X-ray source number
as in Table \ref{xdet}, V, B--V, V--I, match distance between X and optical
position, flux and X-ray luminosity in 0.3-7.9 keV, calculated by assuming 
them as cluster members.}
\begin{tabular}{l l l l l l l l l l l}\\ \hline \hline
Opt id & X Src & V & B--V & V--I &  r & $\log$ f$_\mathrm X$ & $\log$ L$_\mathrm X$ \\
JTH\# &  & mag & mag & mag &  $\arcsec$ &  erg s$^{-1}$ cm$^{-2}$ & erg s$^{-1}$ \\ \hline
5337  &  22  &  16.412  &  1.288  &  1.887  &  0.8  &  -13.44  &  29.81 \\
7417  &  53  &  18.418  &  1.276  &  2.53  &  0.9  &  -13.96  &  29.29 \\
6220  &  67  &  15.123  &  1.177  &  1.481  &  0.9  &  -13.39  &  29.87 \\
9976  &  72  &  17.123  &  1.312  &  1.884  &  0.6  &  -14.04  &  29.21 \\
7485  &  75  &  11.801  &  0.771  &  0.819  &  1.1  &  -13.97  &  29.28 \\
9022  &  76  &  18.448  &  1.74  &  2.587  &  1.6  &  -14.08  &  29.18 \\
8692  &  88  &  18.029  &  1.379  &  2.208  &  0.8  &  -14.68  &  28.57 \\
7200  &  93  &  18.599  &    --    &  2.27  &  2.2  &  -14.48  &  28.78 \\
7708  &  110  &  17.643  &  1.714  &  2.512  &  2.8  &  -13.84  &  29.41 \\
5447  &  167  &  16.555  &  1.348  &  1.912  &  1.5  &  -13.76  &  29.5 \\
7129  &  170  &  18.098  &    --    &  2.101  &  0.2  &  -14.06  &  29.19 \\
7147  &  217  &  13.747  &  0.726  &  0.856  &  0.7  &  -13.55  &  29.7 \\
9492  &  243  &  10.518  &  0.148  &  0.191  &  1.6  &  -14.69  &  28.57 \\
11455  &  264  &  18.759  &  1.348  &  2.646  &  1.5  &  -14.64  &  28.61 \\
11366  &  292  &  12.898  &  0.827  &  0.956  &  2.2  &  -13.07  &  30.18 \\
7844  &  324  &  12.14  &  0.368  &  0.873  &  2  &  -14.05  &  29.21 \\
9106  &  341  &  10.262  &  0.405  &  0.49  &  1.8  &  -14.49  &  28.76 \\
9467  &  348  &  16.95  &  1.132  &  1.825  &  0.3  &  -14.21  &  29.04 \\
4181  &  352  &  20.553  &    --    &  2.865  &  2.6  &  -13.45  &  29.8 \\
6700  &  413  &  12.792  &  0.597  &  0.704  &  2.2  &  -14.08  &  29.17 \\

\hline
\end{tabular}
\end{table*}

A few X-ray sources do not have counterparts in optical or infrared bands. 
After having searched for optical counterparts of the 
X-ray sources in GSC-II, 2MASS and DENIS and the complete catalog
of \citet{Jeffries01}, 49 sources are left without optical/infrared 
known counterparts. 
We have marked these sources with a {\tt U} letter in the last column of 
\ref{xdet} table, the finding charts at their positions are reported in Fig. 
\ref{chart1} and \ref{chart2}.

\section{Summary}
We have presented results from a deep X-ray survey on the young open cluster
NGC~2516, obtained from a series of six XMM-Newton observations. 
The source detection has been carried on the sum of EPIC MOS and pn images 
with a wavelet transform code (PWXDetect).
We have reached fluxes lower by a factor 5 or more with respect to Chandra 
observations carried in the same epoch. 
We have detected 431 sources, 234 of them have as optical counterparts
239 cluster members, five X-ray sources match as many close pairs. 
X-ray emission of spectral type from early B down to M-type stars
has been detected; the coolest detected stars have spectral type M5 and 
T$_{eff} \sim 3000$. 
We have investigated the X-ray spectral properties of cluster stars through
1-T and 2-T models to estimate the dominant plasma thermal components and emission 
measures.  For 1-T model we find temperatures around 0.6 keV. 
In 2-T models the temperatures are in 0.3--0.7 keV and 1.0--2.0 keV ranges
in agreement with the temperatures found in nearly coeval open clusters like Pleiades and 
Blanco 1. 

For each spectral type F, G, K, M we estimated the maximum likelihood 
distribution of X-ray luminosities taking also into account the upper limits 
for undetected cluster stars. 
G-,K-, and M- type stars are under-luminous with 
respect to the Pleiades in the same spectral range. 
Possible biases due to contaminating field stars are discussed and statistical
corrections have been made to X-ray distribution functions. We conclude that
the NGC 2516 solar type stars are definitively less luminous in X-rays than 
the analog Pleiades.
The differences could be attributed mainly to the slight older age of NGC 2516 
(140 Myr vs. 80-100 Myr of Pleiades), less probably to a lower rotation rate.
The past cluster history could also have a role in determining X-ray emission.

By comparing XMM-EPIC and ROSAT-PSPC data we explored X-ray variability on 
time scales comparable to the solar cycle. 
We observe only variations in X-ray luminosities by a factor 2--3. Along with 
what evidenced in other coeval open clusters, this result strongly suggests 
that activity cycles of amplitude and periods like the solar one are not 
present in young Main Sequence stars at an age of $\sim$140 Myr.

We identify 20 likely new candidate members on the basis of their X-ray 
emission and optical photometry. 49 X-ray sources are left unidentified in 
optical and infrared bands and are likely of extra galactic nature and
for them we provide finding charts.

\begin{acknowledgements}
Authors acknowledge financial support by MIUR-PR/N grant
(Ministero dell'Istruzione, Universit\`a e Ricerca).
\end{acknowledgements}
\bibliographystyle{aa}
\bibliography{bibtesi}
\clearpage
\onecolumn
\appendix
\section{Online. X-ray detections}
\renewcommand{\textheight}{25.0cm}
\begin{longtable}{l l l l l l l l}
\caption[]{\label{xdet} List of detected X-ray sources in the combined images of 
XMM-Newton EPIC camera. Columns report: running number of X-ray detection, coordinates,
count rate in MOS1 equivalent counts, detection significance 
(in units of background mean standard deviation), total exposure time, 
offaxis distance and unidentified source flag (cf. Sect. 4).}\\
\hline\hline
Num  & R.A. & Dec. & Count Rate & Signif. & Exp. Time & Offaxis & Un.  \\
 & J2000 & J2000 & cts ks$^{-1}$ & & ks & arcmin &  \\ \hline
\endfirsthead 
\caption[]{continued.}\\
\hline\hline
Num  & R.A. & Dec. & Count Rate & Signif. & Exp. Time & Offaxis  & Un.  \\
 & J2000 & J2000 & cts ks$^{-1}$ & & ks & arcmin  & \\ \hline
\endhead
\hline
\endfoot
\endlastfoot
1 & 07:57:20.63 & -61:4:19.27 & 4.38 $\pm$ 0.82 & 10.74 & 8 & 18.46  &  \\
4 & 07:58:23.5 & -61:3:44.57 & 1.49 $\pm$ 0.27 & 7.59 & 49.55 & 16.56  &  \\
6 & 07:57:17.67 & -61:3:30.6 & 7.05 $\pm$ 0.75 & 14.34 & 16.66 & 17.85  &  \\
7 & 07:57:40.35 & -61:3:31 & 0.62 $\pm$ 0.16 & 5.81 & 46.76 & 16.93  &  \\
8 & 07:58:34.76 & -61:3:17.86 & 1.93 $\pm$ 0.29 & 9.28 & 50.52 & 16.23  &  \\
9 & 07:57:47.6 & -61:3:6.98 & 0.84 $\pm$ 0.17 & 7.21 & 50.04 & 16.32  &  \\
10 & 07:58:26.91 & -61:2:46.14 & 1.16 $\pm$ 0.21 & 8.07 & 54.34 & 15.61  &  \\
11 & 07:58:44.52 & -61:2:37.93 & 0.68 $\pm$ 0.2 & 5.66 & 28.18 & 15.78  &  \\
\end{longtable}

\section{Online. Identifications, X-ray fluxes and luminosities of 
NGC2516 stars}
\begin{longtable}{l l l l l l l l l l l}
\caption[]{\label{xflu} Identifications, X-ray fluxes and luminosities of 
NGC 2516 X-ray detected stars. 
Columns report: X source identifier (as in Table \ref{xdet}, identifier
from \protect{\citet{Jeffries01}}, match distance, V, B--V, V--I, fluxes, luminosities
in the 0.3--7.9 keV band, $\log$ L$_\mathrm{bol}$ and the ratio \lxlbol . 
Errors are propagated from count rates, for any systematic error see discussion
in Sect. \ref{err}. Exposure times are calculated from the summed exposures, 
scaling pn to MOS with a factor 3.6 as described in the text.}\\
\hline\hline
X-ray Src  & Opt. Id. & Distance & V & B--V & V--I & f$_\mathrm X$ (0.3 -- 7.9 keV)& $\log$ L$_\mathrm X$ & $\log$ L$_\mathrm{bol}$ & \lxlbol \\ 
 & & arcsec & mag & mag & mag & $10^{-15}\cdot$ erg s$^{-1}$ cm$^{-2}$ & erg s$^{-1}$ & erg s$^{-1}$ &  \\ \hline
\endfirsthead 
\caption[]{continued.}\\
\hline\hline
X-ray Src  & Opt. Id. & Distance & V & B--V & V--I & f$_\mathrm X$ (0.3 -- 7.9 keV)& $\log$ L$_\mathrm X$ & $\log$ L$_\mathrm{bol}$ & \lxlbol \\ 
 & & arcsec & mag & mag & mag & $10^{-15}\cdot$ erg s$^{-1}$ cm$^{-2}$ & erg s$^{-1}$ & erg s$^{-1}$ &  \\ \hline
\endhead
\hline
\endfoot
\endlastfoot
6  &  6160  &  1  &  20.9  &  ...  &  3.06  &  67 $\pm$ 7  &  30.08$_{-0.05}^{0.04}$ & 31.52 & -1.44 \\
7  &  7036  &  3.8  &  13  &  0.7  &  0.81  &  6 $\pm$ 2  &  29.02$_{-0.1}^{0.1}$ & 33.9 & -4.88 \\
8  &  9114  &  5.9  &  13.1  &  0.7  &  0.8  &  18 $\pm$ 3  &  29.52$_{-0.07}^{0.06}$ & 33.93 & -4.41 \\
10  &  8793  &  2.3  &  20.1  &  ...  &  3.02  &  11 $\pm$ 2  &  29.29$_{-0.08}^{0.07}$ & 31.54 & -2.24 \\
11  &  9544  &  3.5  &  17.5  &  1.5  &  2.25  &  6.5 $\pm$ 2  &  29.07$_{-0.2}^{0.1}$ & 31.94 & -2.87 \\
12  &  6976  &  4.3  &  12.8  &  0.6  &  0.78  &  46 $\pm$ 4  &  29.91$_{-0.04}^{0.03}$ & 33.97 & -4.05 \\
15  &  6871  &  5.4  &  14.8  &  1.2  &  1.34  &  7 $\pm$ 2  &  29.1$_{-0.1}^{0.1}$ & 32.93 & -3.83 \\
16  &  7682  &  1.3  &  19.4  &  ...  &  2.64  &  10 $\pm$ 2  &  29.25$_{-0.09}^{0.07}$ & 31.68 & -2.43 \\
\hline
\end{longtable}
\section{Online. X-ray upper limits of NGC2516 stars}
\begin{longtable}{l l l l l l l l l l l}
\caption[]{\label{xupl} X-ray undetected stars of NGC 2516. 
Columns report: identifier from \protect{\citet{Jeffries01}}, V, B--V, V--I, exposure
time (ks), upper limit to rate (ct/ks),  flux and luminosities in the 0.3--7.9 
keV band, $\log$ L$_\mathrm{bol}$ and the ratio \lxlbol . }\\
\hline\hline
Opt. Id. &  V & B--V & V--I & Time & Limit Rate  & Lim. $\log f_\mathrm X$ & Lim. $\log$ L$_\mathrm X$ &$\log$ L$_\mathrm{bol}$ & \lxlbol\\ 
 & mag & mag & mag & ks & ct ks$^{-1}$ &erg s$^{-1}$ cm$^{-2}$ &erg s$^{-1}$ & erg s$^{-1}$ & \\
\hline
\endfirsthead 
\caption[]{continued.}\\
\hline\hline
Opt. Id. &  V & B--V & V--I & Time & Limit Rate  & Lim. $\log f_\mathrm X$ & Lim. $\log$ L$_\mathrm X$ &$\log$ L$_\mathrm{bol}$ & \lxlbol \\ 
 & mag & mag & mag & ks & ct ks$^{-1}$ &erg s$^{-1}$ cm$^{-2}$ &erg s$^{-1}$  & erg s$^{-1}$ & \\
\hline
\endhead
\hline
\endfoot
\endlastfoot
3680  &  17.238  &  1.537  &  1.909  &  11.28  &  $\leq$ 1.321  &  $\leq$-13.9  &  $\leq$29.35  &  32.34  &  $\leq$-2.99 \\
3748  &  15.753  &  1.211  &  1.446  &  11.28  &  1.443  &  -13.86  &  29.39  &  32.84  &  -3.45 \\
3777  &  20.005  &  ...  &  3.113  &  11.34  &  1.439  &  -13.86  &  29.39  &  31.51  &  -2.12 \\
3891  &  17.183  &  1.564  &  2.022  &  11.49  &  1.324  &  -13.9  &  29.35  &  32.18  &  -2.83 \\
4069  &  19.661  &  ...  &  2.964  &  11.73  &  1.35  &  -13.89  &  29.36  &  31.55  &  -2.19 \\
4072  &  18.981  &  ...  &  2.47  &  12.14  &  1.253  &  -13.92  &  29.33  &  31.76  &  -2.43 \\
4146  &  20.693  &  ...  &  3.361  &  27.16  &  1.077  &  -13.99  &  29.26  &  31.42  &  -2.16 \\
4187  &  14.492  &  1.112  &  1.214  &  23.01  &  0.87  &  -14.08  &  29.17  &  33.08  &  -3.91 \\
\hline
\end{longtable}
\vfill
\section{Online. Finding charts of unidentified sources}
\begin{figure*}
\includegraphics[width=18cm]{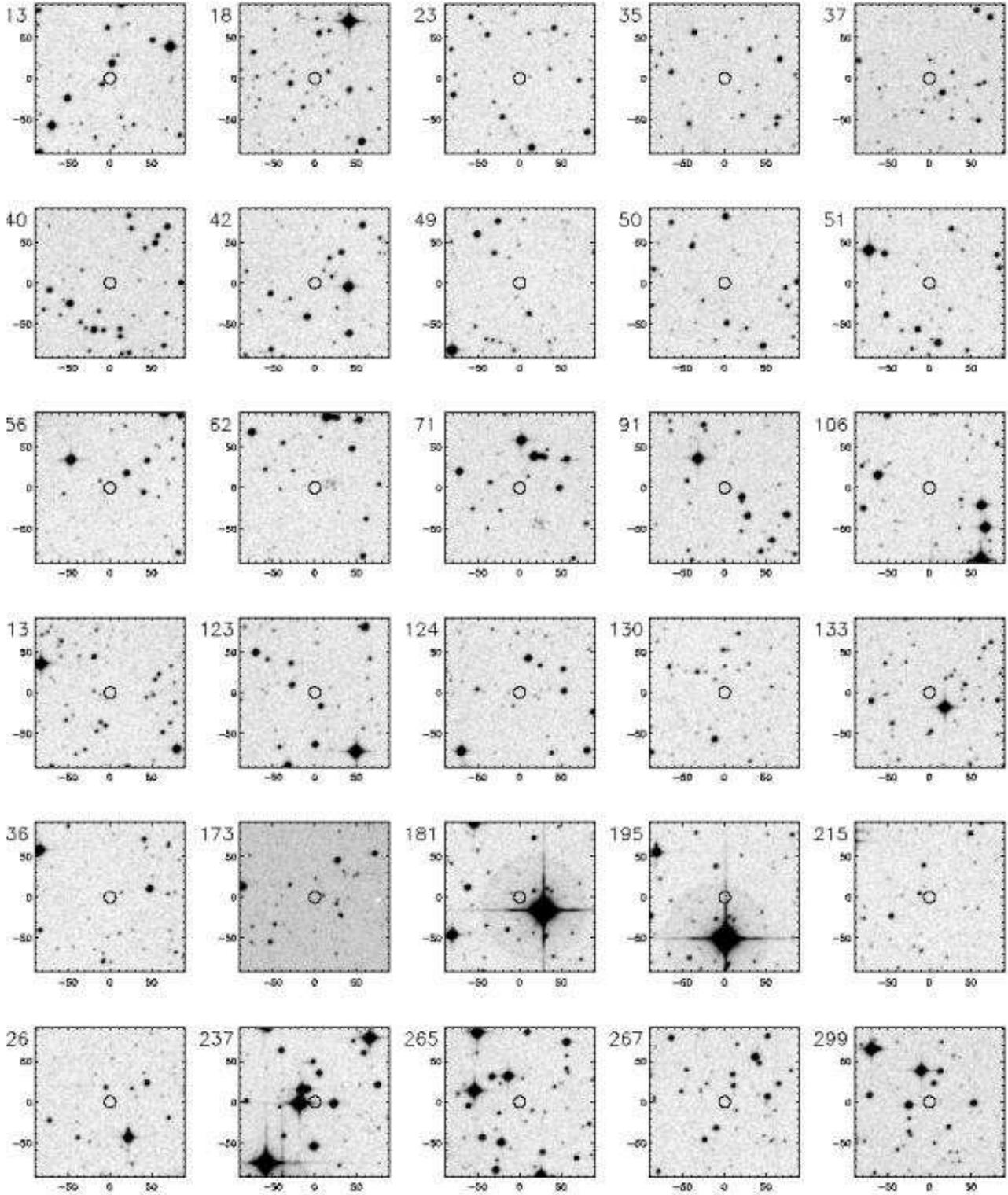}
\caption{\label{chart1}
DSS IR finding charts ($3^\prime \times 3^\prime $ side) 
in the position of unidentified X-ray sources}
\end{figure*}
\newpage
\begin{figure*}
\includegraphics[width=18cm]{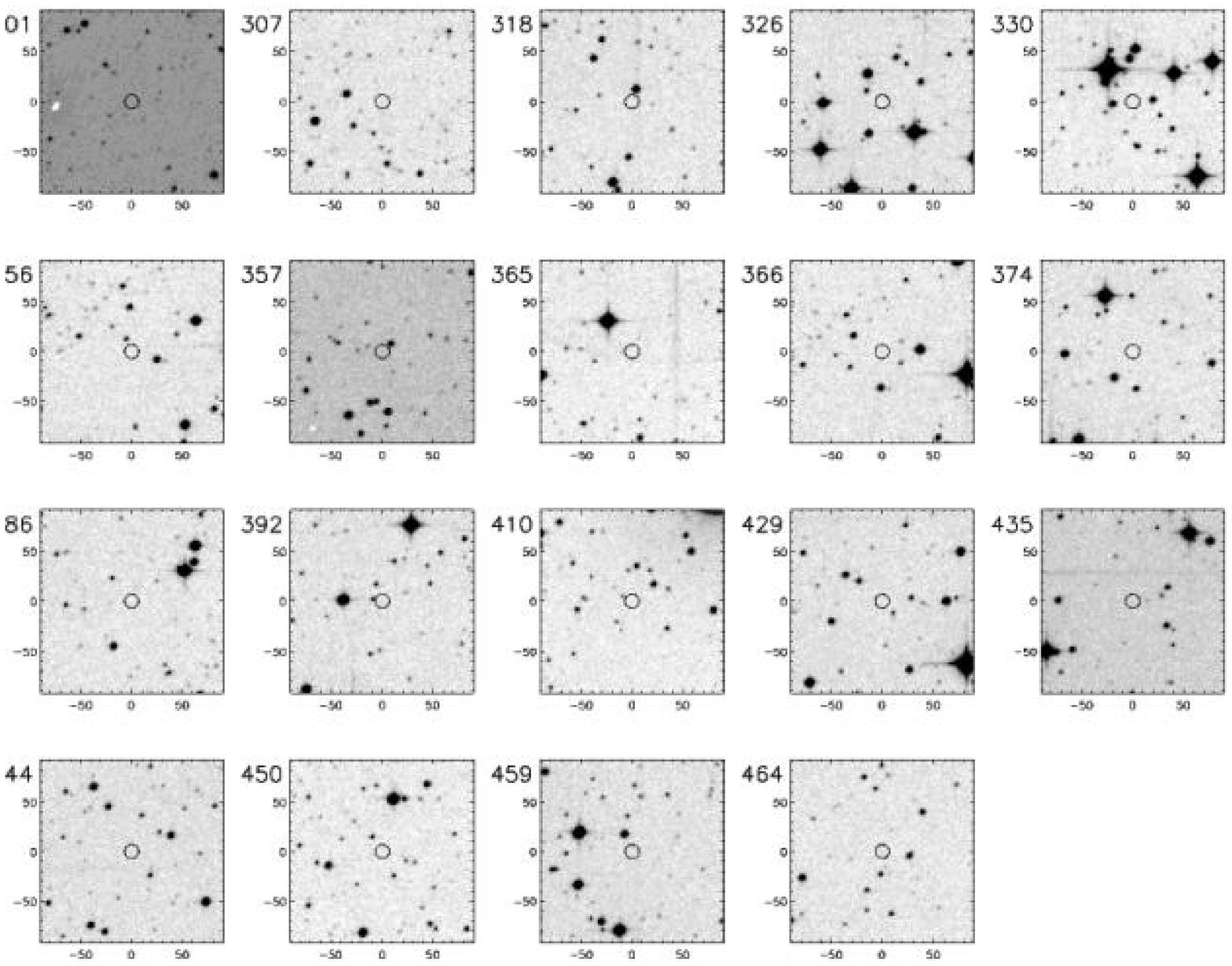}
\caption{\label{chart2} As in Fig. \ref{chart1} }
\end{figure*}
\end{document}